\def\fracm#1#2{\hbox{\large{${\frac{{#1}}{{#2}}}$}}}
\def\dvp{\raisebox{-.45ex}{\rlap{$
=$}} \raisebox{-.45ex}{$\hskip .48ex { |}$}}
\def\DP{{\scriptsize{\dvp}}~~}
\documentstyle[12pt]{article}

\skewchar\fivmi='177
\skewchar\sixmi='177
\skewchar\sevmi='177
\skewchar\egtmi='177
\skewchar\ninmi='177
\skewchar\tenmi='177
\skewchar\elvmi='177
\skewchar\twlmi='177
\skewchar\frtnmi='177
\skewchar\svtnmi='177
\skewchar\twtymi='177
\def\@magscale#1{ scaled \magstep #1}
\skewchar\fivsy='60
\skewchar\sixsy='60
\skewchar\sevsy='60
\skewchar\egtsy='60
\skewchar\ninsy='60
\skewchar\tensy='60
\skewchar\elvsy='60
\skewchar\twlsy='60
\skewchar\frtnsy='60
\skewchar\svtnsy='60
\skewchar\twtysy='60

\catcode`@=11
\def\un#1{\relax\ifmmode\@@underline#1\else
        $\@@underline{\hbox{#1}}$\relax\fi}
\catcode`@=12

\def\a{\alpha}
\def\b{\beta}

\def\d{\delta}
\def\e{\epsilon}

\def\g{\gamma}

\def\l{\lambda}

\def\s{\sigma}
\def\t{\tau}

\def\L{\Lambda}

\def\dslash{\not{\hbox{\kern-2pt $\partial$}}}
\def\Dslash{\not{\hbox{\kern-4pt $D$}}}
\def\pslash{\not{\hbox{\kern-2.3pt $p$}}}
 \newtoks\slashfraction
 \slashfraction={.13}
 \def\slash#1{\setbox0\hbox{$ #1 $}
 \setbox0\hbox to \the\slashfraction\wd0{\hss \box0}/\box0 }

\font\ro=cmsy10                          
\def\kcr{{\hbox{\ro \char'170}}}                
\def\ktl{{\hbox{\ro \char'170}}}        
\def\ktr{{\hbox{\ro \char'170}}}        
\def\kbl{{\hbox{\ro \char'170}}}        
\def\kbr{{\hbox{\ro \char'170}}}        

\def\plpl{\raise-2pt\hbox{$\raise3pt\hbox{$_+$}\hskip-6.67pt\raise0.0pt
\hbox{$^+$}\hskip 0.01pt$}}
\def\mimi{\raise-2pt\hbox{$\raise3pt\hbox{$_-$}\hskip-6.67pt\raise0.0pt
\hbox{$^-$}\hskip 0.01pt$}}

\def\bo{{\raise.15ex\hbox{\large$\Box$}}}               
\def\pa{\partial}                                       
\def\TH{{\raise.2ex\hbox{$\displaystyle \bigodot$}\mskip-4.7mu \llap H \;}}
\def\face{{\raise.2ex\hbox{$\displaystyle \bigodot$}\mskip-2.2mu \llap {$\ddot
        \smile$}}}                                      

\def\sp#1{{}^{#1}}                              
\def\sb#1{{}_{#1}}                              
\def\leftrightarrowfill{$\mathsurround=0pt \mathord\leftarrow \mkern-6mu
        \cleaders\hbox{$\mkern-2mu \mathord- \mkern-2mu$}\hfill
        \mkern-6mu \mathord\rightarrow$}
\def\dvec#1{\vbox{\ialign{##\crcr
        \leftrightarrowfill\crcr\noalign{\kern-1pt\nointerlineskip}
        $\hfil\displaystyle{#1}\hfil$\crcr}}}           

\def\fracm#1#2{\hbox{\large{${\frac{{#1}}{{#2}}}$}}}
\def\frac#1#2{{\textstyle{#1\over\vphantom2\smash{\raise.20ex
        \hbox{$\scriptstyle{#2}$}}}}}                   
\def\sfrac#1#2{{\vphantom1\smash{\lower.5ex\hbox{\small$#1$}}\over
        \vphantom1\smash{\raise.4ex\hbox{\small$#2$}}}} 
\def\bfrac#1#2{{\vphantom1\smash{\lower.5ex\hbox{$#1$}}\over
        \vphantom1\smash{\raise.3ex\hbox{$#2$}}}}       
\def\afrac#1#2{{\vphantom1\smash{\lower.5ex\hbox{$#1$}}\over#2}}    

\newskip\humongous \humongous=0pt plus 1000pt minus 1000pt

\newif\ifdtup

\def\ref#1{$\sp{#1)}$}

\parskip=\medskipamount                 
\thicklines                         

\def\oldheadpic{                                
        \setlength{\unitlength}{.4mm}
        \thinlines
        \par
        \begin{picture}(349,16)
        \put(325,16){\line(1,0){4}}
        \put(330,16){\line(1,0){4}}
        \put(340,16){\line(1,0){4}}
        \put(335,0){\line(1,0){4}}
        \put(340,0){\line(1,0){4}}
        \put(345,0){\line(1,0){4}}
        \put(329,0){\line(0,1){16}}
        \put(330,0){\line(0,1){16}}
        \put(339,0){\line(0,1){16}}
        \put(340,0){\line(0,1){16}}
        \put(344,0){\line(0,1){16}}
        \put(345,0){\line(0,1){16}}
        \put(329,16){\oval(8,32)[bl]}
        \put(330,16){\oval(8,32)[br]}
        \put(339,0){\oval(8,32)[tl]}
        \put(345,0){\oval(8,32)[tr]}
        \end{picture}
        \par
        \thicklines
        \vskip.2in}
\def\oldtitle#1#2#3#4{\oldheadpic\begin{center}\vglue.5in{\large\bf #1}\\[.6in]
        {#2}\\[.1in] {\it Department of Physics and Astronomy}\\
        {\it University of Maryland, College Park, MD 20742}\\[.6in]
        Physics Publication \#{#3}\\ {#4}\\[1.5in] {\bf ABSTRACT}\\[.1in]
        \end{center} \begin{quotation}}                 
\def\oldTitle#1#2#3#4#5#6#7{\oldheadpic\begin{center} \vglue .4in
        {\large\bf #1}\\[.4in]
        {#2}\\[.1in] {\it Department of Physics and Astronomy}\\
        {\it University of Maryland, College Park, MD 20742}\\[.1in]
        {#3}\\[.1in] {\it {#4}}\\ {\it {#5}}\\[.4in]
        Physics Publication \#{#6}\\ {#7}\\[.5in] {\bf ABSTRACT}\\[.1in]
        \end{center} \begin{quotation}}                 
\def\border{                                            
        \setlength{\unitlength}{1mm}
        \newcount\xco
        \newcount\yco
        \xco=-21
        \yco=12
        \begin{picture}(140,0)
        \put(\xco,\yco){$\ktl$}
        \advance\yco by-1
        {\loop
        \put(\xco,\yco){$\kcr$}
        \advance\yco by-2
        \ifnum\yco>-240
        \repeat
        \put(\xco,\yco){$\kbl$}}
        \xco=158
        \yco=12
        \put(\xco,\yco){$\ktr$}
        \advance\yco by-1
        {\loop
        \put(\xco,\yco){$\kcr$}
        \advance\yco by-2
        \ifnum\yco>-240
        \repeat
        \put(\xco,\yco){$\kbr$}}
        \put(-20,13){\tiny University of Maryland Elementary Particle
Physics University of Maryland Elementary Particle Physics University of
Maryland Elementary Particle Physics}
        \put(-20,-241.5){\tiny University of Maryland Elementary
Particle Physics University of Maryland Elementary Particle Physics
University of Maryland Elementary Particle Physics}
        \end{picture}
        \par\vskip-8mm}
\def\bordero{                                           
        \setlength{\unitlength}{1mm}
        \newcount\xco
        \newcount\yco
        \xco=-31
        \yco=12
        \begin{picture}(140,0)
        \put(\xco,\yco){$\ktl$}
        \advance\yco by-1
        {\loop
        \put(\xco,\yco){$\kclr}
        \advance\yco by-2
        \ifnum\yco>-240
        \repeat
        \put(\xco,\yco){$\kbl$}}
        \xco=151
        \yco=12
        \put(\xco,\yco){$\ktr$}
        \advance\yco by-1
        {\loop
        \put(\xco,\yco){$\kcr$}
        \advance\yco by-2
        \ifnum\yco>-240
        \repeat
        \put(\xco,\yco){$\kbr$}}
        \put(-20,12){\ooo
bacdefghidfghghdhededbihdgdfdfhhdheidhdhebaaahjhhdahba

hgdedge
   hgfdiehhgdigicba}
        \put(-20,-241.5){\ooo
ababaighefdbfghgeahgdfgafagihdidihiidhiagfedhadbfd

ecdcdfa
   gdcbhaddhbgfchbgfdacfediacbabab}
        \end{picture}
        \par\vskip-8mm}
\def\headpic{                                           
        \indent
        \setlength{\unitlength}{.4mm}
        \thinlines
        \par
        \begin{picture}(29,16)
        \put(165,16){\line(1,0){4}}
        \put(170,16){\line(1,0){4}}
        \put(180,16){\line(1,0){4}}
        \put(175,0){\line(1,0){4}}
        \put(180,0){\line(1,0){4}}
        \put(185,0){\line(1,0){4}}
        \put(169,0){\line(0,1){16}}
        \put(170,0){\line(0,1){16}}
        \put(179,0){\line(0,1){16}}
        \put(180,0){\line(0,1){16}}
        \put(184,0){\line(0,1){16}}
        \put(185,0){\line(0,1){16}}
        \put(169,16){\oval(8,32)[bl]}
        \put(170,16){\oval(8,32)[br]}
        \put(179,0){\oval(8,32)[tl]}
        \put(185,0){\oval(8,32)[tr]}
        \end{picture}
        \par\vskip-6.5mm
        \thicklines}
\def\title#1#2#3#4{\border\headpic {\hbox to\hsize{#4 \hfill UMDEPP #3}}\par
        \begin{center} \vglue .5in {\large\bf #1}\\[.6in]
        {#2}\\[.1in] {\it Department of Physics and Astronomy}\\
        {\it University of Maryland, College Park, MD 20742}\\[1.5in]
        {\bf ABSTRACT}\\[.1in] \end{center} \begin{quotation}}  
\def\Title#1#2#3#4#5#6#7{\border\headpic
        {\hbox to\hsize{#7 \hfill UMDEPP #6}}\par
        \begin{center} \vglue .4in {\large\bf #1}\\[.4in]
        {#2}\\[.1in] {\it Department of Physics and Astronomy}\\
        {\it University of Maryland, College Park, MD 20742}\\[.1in]
        {#3}\\[.1in] {\it {#4}}\\ {\it {#5}}\\[.5in] {\bf ABSTRACT}\\[.1in]
        \end{center} \begin{quotation}}                 
\def\endtitle{\end{quotation}\newpage}                  

\begin{document}

\def\gfrac#1#2{\frac {\scriptstyle{#1}}
        {\mbox{\raisebox{-.6ex}{$\scriptstyle{#2}$}}}}
\def\gg{{\hbox{\sc g}}}
\border\headpic {\hbox to\hsize{
September 1994 \hfill {UMDEPP 95-49}}}
\par
\setlength{\oddsidemargin}{0.3in}
\setlength{\evensidemargin}{-0.3in}
\begin{center}
\vglue .08in
{\large\bf Ultra-Multiplets: A New
Representation of \\ Rigid 2D, N = 8 Supersymmetry}
\\[.72in]

S. James Gates, Jr.\footnote {Supported in part by National Science
Foundation Grant
PHY-91-19746}
${}^,$ \footnote {Supported in part by NATO Grant CRG-93-0789}
and L. Rana
\\[.02in]
{\it Department of Physics\\
University of Maryland\\
College Park, MD 20742-4111  USA}\\
{\bf {\tt gates@umdhep.umd.edu}}\\[1.2in]

{\bf ABSTRACT}\\[.002in]
\end{center}
\begin{quotation}
{By utilizing a new procedure (the RADIO method) for deriving on-shell
2D, 2N-extended multiplets from off-shell 2D, N-extended multiplets,
we derive a new on-shell 2D, N = 8 representation; the ultra-multiplet.
By twisting with respect to parity, we show that many variant versions
of this supermultiplet also exist.}

\endtitle
\section{Introduction}

{}~~~~There is little general understanding of the systematics of irreducible
representations of supersymmetry.  This is reflected in the fact that
most of the recognized on-shell representations (i.e. the supersymmetry
algebra is satisfied only with the use of the equations of motions) do
not presently have off-shell representations (i.e. the supersymmetry
algebra is satisfied without the use of the equations of motions). An
off-shell representation of supersymmetry is obtained when a complete
set of auxiliary fields is added to the on-shell theory. With the re-birth
of string theory almost a decade ago \cite{GS,Het}, it was possible to
hope that a resolution to this problem might be forthcoming from the
study of superstrings since these theories also require auxiliary fields
for their complete description.  Unfortunately, superstring theory has
effectively contributed little to the resolution of our problem. This is
not a criticism of superstring theory. Instead it is a reflection of how
poor is our understanding of superstring theory.

{}~~~~So we are thrown back to artifice, diligence, fortune and insight
to make further progress on the off-shell supersymmetric representation
problem.  A closely related problem is that of finding explicit
irreducible representations for large values of N, the degree of
``extendedness'' of the supersymmetry. An important place to study these
problems is within the realm of a two-dimensional space-time. This is
an interesting realm in which to explore this question because solutions
have consequences back on superstring theory, integrable systems and
mathematical profundities.  In 1988 \cite{GaLO}, progress was made on
the case of the N = 4 theory with the discovery of a minimal off-shell 2D,
N = 4 supergravity multiplet.   More recently there has been a
clarification of the situation with N = 4 superstrings \cite{Ga}, where
it was demonstrated that at least three off-shell 2D, N = 4 superstrings
actions exist and that likely one more such theory should be
possible.\footnote{Actually, we now know of a total of 7 possibly
distinct N = 4 superstring actions!}

\section{RADIO: A ``Chemical''
Derivation of a New Supersymmetric Representation}

{}~~~~Before we present the explicit realization of 2D, N = 8 supersymmetry
provided by the ultra-multiplet, it is useful possibly for future research
to explain the method by which the ultra-multiplet was found.  The genesis
of our discovery is a very interesting process that we shall call
``reduction, automorphic dualization, integration \& oxidation'' (RADIO).
We will not provide a general proof of why this method works.
Instead we will simply use it.

{}~~~~Our starting point begins with off-shell 2D, N = 4 twisted
hypermultiplets
(THM-I and THM-II)\footnote{These were called TM-I and TM-II in reference
\cite{Ga} and are described by equations (1) and (4) there.}.  There are
different reductions possible to 1D.  Let us concentrate on the reduction
of the THM-II model, where we consider a simple toroidal compactification with
all fields only dependent on the $\t$-coordinate of the world sheet.  However,
we retain all of the Grassmann coordinates of the original 2D theory. This
has the effect of doubling the number of supersymmetries. So we go from a
2D, N = 4 model to a 1D, N = 8 model.  At this stage, equation (4) of
reference \cite{Ga} reads,
$$~~~~~~~~~~~~~{D}_{I i } {\cal T}  ~=~
(\g \sp{3})  \sb{I} \sp{J} {\Psi}_{J i}  ~~~,
{}~~~~~~~~~~~~~~~~~~~~~~~~~~~~~~~~~~~~~~~~~~~~~~~~~~$$
$$~~~~~~~~~~~~~~~~~~~~~{D}_{I i } {\cal X} {}_j {}^k ~=~   i \left [
\d_i {}^k { \Psi}_{I j} ~-~ \fracm 12 \d_j {}^k {\Psi}_{I i}  \right ] ~~~,
{}~~~~~~~~~~~~~~~~~~~~~~~~~~~~~~~~~~~~~~~~~~~~~~~~$$
$$~~~~~~~~~~~~~~~~~~~~~~~~~~~~~~ {\cal X} {}_i {}^i  ~=~ 0 ~~~, ~~~ {\cal X}
{}_i {}^j ~-~ ( {\cal X} {}_j {}^i )^* ~=~ 0 ~~~,
{}~~~~~~~~~~~~~~~~~~~~~~~~~~~~~~~~~~~~~~~~~~~~~~~~~~~$$
$$~~~~~~~~~~~~~{D}_{I i } {\Psi}_{J j}  ~=~ \fracm 12  C_{ i j} C_{I J}
{\bar J}~~~, ~~~~~~~~~~~~~~~~~~~~~~~~~~~~~~~~~~~~~~~~~~~~~~~~~~~~~~$$
$$~~~~~~~~~~~~~~~~~~~~~~~~~~~~{D}_{\a i } {\bar J} ~=~ 0 ~~~,~~~ m ~-~
({m} )^* ~=~ 0 ~~~, n ~-~ ({n} )^* ~=~ 0 ~~~, ~~~~~~~~~~~~~~~~~~~~~~~~~~~~$$
$${\bar D}^{I i } {\Psi}_{J j}  ~=~ i \d_j {}^i
(\g \sp{3} \g \sp{0})  \sb{J} \sp{I}  \left ( \pa_{\t} {\cal T} \right ) ~+~ 2
(\g \sp{0})  \sb{J} \sp{I} \left ( \pa_{\t} \sb{a} {\cal X}_{j}{}^i \right )
{}~~~~~~~$$
$$~~~~~~~~~~~~~
 ~+~ i \fracm 12 \d_{J} \sp{I} \d \sb j {}^i m ~+~ \fracm 12 (\g \sp{3})
\sb{J} \sp{I}  \d \sb j {}^i n ~~. ~~~~~~~~~~$$
$$ ~~~~~{D}_{I i } { J} ~=~  i 4  C_{ i j} (\g \sp{0})  \sb{I J}
\left ( \pa_{\t} {\bar \Psi}^{J j} \right ) ~~~,~~~~~~~~~~~~~~
{}~~~~~~~~~~~ $$
$$ ~~~{D}_{I i } n ~=~ - i 2 (\g \sp{3} \g \sp{0})  \sb{I} \sp{J}
\left ( \pa_{\t}  { \Psi}_{J i} \right ) ~~~,~~~~~~~~~~~~~~~~~~~
{}~~~~~~~~ $$
\begin{equation}
{}~{D}_{I i } m ~=~ - 2 ( \g \sp{0})  \sb{I} \sp{J}
\left ( \pa_{\t}  { \Psi}_{J i} \right ) ~~~,~~~~~~~~~~~~~~~~
{}~~~~~~~~~~~~~
\end{equation}
where we have replaced all spinor indices ($\a$, $\b$,...) by
internal symmetry indices (i.e. $I$, $J$,...) to emphasize their
no longer being related to spin.

{}~~~~Next we perform a transformation that we call a ``1D automorphic
duality transformation'' \cite{GRana}.  Our investigation within the
realm of 1D supersymmetric representations seem to indicate that
auxiliary fields can be avoided entirely in 1D.  This is a very
unusual ``transform'' that formally replaces the would-be ``auxiliary
fields'' of a 1D supermultiplet by propagating fields. What this
 amounts to is replacing every would-be ``auxiliary field'' by
the $\t$-derivative of a new field.  When this is done, it can be
observed that $\cal T$, $m$ and $n$ form a SU(2)-triplet in the
space of the $I$-indices.  With a little bit of redefinition of
fields, we obtain our final result that defines the 1D ultra-multiplet.
We can dispense with the equations above since their only use was
to help us derive the final result below defining the 1D
ultra-multiplet. Explicitly, its supersymmetry variations take
the forms,
$$~~~~~~~~~~~~~{D}_{I i } {\cal G}  ~=~  i 2  C_{i j}  C_{I J}
{\bar {\varphi}}^{J j}  ~~~,
{}~~~~~~~~~~~~~~~~~~~~~~~~~~~~~~~~~~~~~~~~~~~~~~~~~~$$
$$~~~~{\bar D}^{I i } {\cal G}  ~=~ 0 ~~~, ~~~~~~
{}~~~~~~~~~~~~~~~~~~~~~~~~~~~~~~~~~~~~~~~~~~~~~~~~~~$$
$$~~~~~~~~~~~~~~~~~~~~~{D}_{I i } {L} {}_j {}^k ~=~   i 2 \left [
\d_i {}^k { \varphi }_{I j} ~-~ \fracm 12 \d_j {}^k {\varphi}_{I i}
\right ] ~~~,~~~~
{}~~~~~~~~~~~~~~~~~~~~~~~~~~~~~~~~~~~~~~~~~~~~~~~~$$
$$~~~~~~~~~~~~~~~~~~~~~{D}_{I i } {R} {}_J {}^K ~=~    2 \left [
\d_I {}^K { \varphi }_{J i} ~-~ \fracm 12 \d_J {}^K {\varphi}_{I i}
\right ] ~~~,~~~~~~~~
{}~~~~~~~~~~~~~~~~~~~~~~~~~~~~~~~~~~~~~~~~~~~~~~~~$$
$$~~~~~~~~~~~~~{D}_{I i } { \varphi }_{J j}  ~=~ -  C_{i j}  C_{I J}
(\pa_{\t} {\bar {\cal G}} ) ~~~,~~~
{}~~~~~~~~~~~~~~~~~~~~~~~~~~~~~~~~~~~~~~~~~~~~~~~~$$
\begin{equation}
{}~~~~{\bar D}^{I i } { \varphi }_{J j}  ~=~ i( \pa_{\t}  {R} {}_J {}^I ) ~
\d_j {}^i ~+~ ( \pa_{\t} {L} {}_j {}^i ) ~ \d_J{}^I ~~~.~~~~~~~~~~~~
{}~~~~~~~~~~~~~ \end{equation}
Thus, the 1D, ultra-multiplet consists of eight bosons ${\cal G}$,
$ {L} {}_j {}^k$ and ${R} {}_J {}^K$ as well as eight fermions
${ \varphi }_{I i}$.

While in 1D, there is second type of multiplet that can be constructed.
To construct it we utilize the 1D automorphic duality transformation
once again.  This  can be done by using the results in (2) as our
starting point.  First there is simply a ``change'' the name of the
spinor ${ \varphi }_{I i} \to { \zeta }_{I i}$.  Next the 1D automorphic
duality transformation is implemented by acting {\underline{only}}
{\underline {on}} {\underline {each}} scalar field transformation
law with $\pa_{\t}$.  After this step, all scalar fields in the
transformation laws appear only through their $\t$ derivatives.
These $\t$ derivative terms are then replaced by new independent
bosonic fields without $\t$ derivatives.  This is also 1D automorphic
duality map.  Carrying out all of these step, we are led to the
``fermionic ultra-multiplet'' (FUM) transformation laws,
$$~~~~~~~~~~~~~{D}_{I i } {\cal C}  ~=~  i 2  C_{i j}  C_{I J}
\pa_{\t} {\bar {\zeta}}^{J j}  ~~~,
{}~~~~~~~~~~~~~~~~~~~~~~~~~~~~~~~~~~~~~~~~~~~~~~~~~~$$
$$~~~~{\bar D}^{I i } {\cal C}  ~=~ 0 ~~~, ~~~~~~~~
{}~~~~~~~~~~~~~~~~~~~~~~~~~~~~~~~~~~~~~~~~~~~~~~~~~~$$
$$~~~~~~~~~~~~~~~~~~~~~{D}_{I i } {\cal C} {}_j {}^k ~=~   i 2 \left [
\d_i {}^k \pa_{\t} { \zeta }_{I j} ~-~ \fracm 12 \d_j {}^k
\pa_{\t} {\zeta}_{I i}   \right ] ~~~,~~~~
{}~~~~~~~~~~~~~~~~~~~~~~~~~~~~~~~~~~~~~~~~~~~~~~~~$$
$$~~~~~~~~~~~~~~~~~~~~~~{D}_{I i } {\cal C} {}_J {}^K ~=~    2 \left [
\d_I {}^K \pa_{\t} { \zeta }_{J i} ~-~ \fracm 12 \d_J {}^K
\pa_{\t} {\zeta}_{I i}
\right ] ~~~,~~~~~~~~~
{}~~~~~~~~~~~~~~~~~~~~~~~~~~~~~~~~~~~~~~~~~~~~~~~~$$
$$~~~~~~~~~~~~~{D}_{I i } { \zeta }_{J j}  ~=~ -  C_{i j}  C_{I J}
 {\bar {\cal C}}  ~~~,~~~~~~~~~~
{}~~~~~~~~~~~~~~~~~~~~~~~~~~~~~~~~~~~~~~~~~~~~~~~~$$
\begin{equation}
{}~~~~{\bar D}^{I i } { \zeta }_{J j}  ~=~ i  {\cal C} {}_J {}^I  ~
\d_j {}^i ~+~  {\cal C} {}_j {}^i  ~ \d_J{}^I ~~~.~~~~~~~~~~~~
{}~~~~~~~~~~~~~~~~~~~~~~~~ \end{equation}

It is a simple matter to show that the supersymmetry variations
above uniformly yield a representation of the 1D supersymmetry algebra;
\begin{equation}
\{ {D}_{I i }~,~ {D}_{J j } \} ~=~ 0 ~~~,~~~  \{ {D}_{I i }~,~
{\bar {D}}^{J j } \} ~=~ i 2 \d_i {}^j  \d_I {}^J \pa_{\t} ~~~.
 \end{equation}
This completes the ``reduction'' procedure of the process. Next we
begin the ``oxidation'' procedure.

{}~~~~Were we to explicitly write out the fermionic derivative in equation (3),
we would find that it depends on one bosonic derivative and 8 Grassmannian
derivatives as well as their associated coordinates.  So the first step
of the ``oxidation'' is to realize that we can consider a transformation
that replaces $\pa_{\t}$ by $\pa_{\DP}$ and thus go up to a 2D heterotic
model with (8,0) supersymmetry!  In appearance it is almost identical
to the equations above with the exception that ``$+$'' indices must be
appropriately inserted into the equations,
$$~~~~~~~~~~~~~{D}_{I i +} {\cal G}  ~=~  i 2  C_{i j}  C_{I J}
{\bar {\varphi}}{}_+ {}^{J j}  ~~~,
{}~~~~~~~~~~~~~~~~~~~~~~~~~~~~~~~~~~~~~~~~~~~~~~~~~~$$
$$~~~~{{\bar D}_{+}}^{I i } {\cal G}  ~=~ 0 ~~~, ~~~~~~
{}~~~~~~~~~~~~~~~~~~~~~~~~~~~~~~~~~~~~~~~~~~~~~~~~~~$$
$$~~~~~~~~~~~~~~~~~~~~~{D}_{I i +} {L} {}_j {}^k ~=~   i 2 \left [
\d_i {}^k { \varphi }_{I j +} ~-~ \fracm 12 \d_j {}^k {\varphi}_{I i +}
\right ] ~~~,~~~~
{}~~~~~~~~~~~~~~~~~~~~~~~~~~~~~~~~~~~~~~~~~~~~~~~~$$
$$~~~~~~~~~~~~~~~~~~~~~{D}_{I i +} {R} {}_J {}^K ~=~    2 \left [
\d_I {}^K { \varphi }_{J i +} ~-~ \fracm 12 \d_J {}^K {\varphi}_{I i +}
\right ] ~~~,~~~~~~~~
{}~~~~~~~~~~~~~~~~~~~~~~~~~~~~~~~~~~~~~~~~~~~~~~~~$$
$$~~~~~~~~~~~~~{D}_{I i +} { \varphi }_{J j +}  ~=~ -  C_{i j}  C_{I J}
(\pa_{\DP} {\bar {\cal G}} ) ~~~,~~~
{}~~~~~~~~~~~~~~~~~~~~~~~~~~~~~~~~~~~~~~~~~~~~~~~~~$$
\begin{equation}
{}~~~~~{\bar D}^{I i +} { \varphi }_{J j +}  ~=~ i( \pa_{\DP}  {R} {}_J {}^I )
{}~
\d_j {}^i ~+~ ( \pa
                                                      _{\DP} {L} {}_j {}^i ) ~
\d_J{}^I ~~~,~~~~~~~~~~~~
{}~~~~~~~~~~~~~ \end{equation}
where now we have a realization of the 2D, (8,0) heterotic supersymmetry
algebra
\begin{equation}
\{ {D}_{I i +}~,~ {D}_{J j +} \} ~=~ 0 ~~~,~~~  \{ {D}_{I i +}~,~
{{\bar D}_+}^{J j } \} ~=~ i 2 \d_i {}^j  \d_I {}^J \pa_{\DP} ~~~.
\end{equation}

{}~~~~The astute reader may well guess what is to follow.  The fermionic
ultra-multiplet can also be oxidized into a (8,0) representation!
As may be guessed from the form of the supersymmetry variations in
(3), the FUM naturally oxidizes into an (8,0) ``minus spinor''
\cite{BGM} multiplet.  We simply need to judiciously introduce
``$-$'' indices into (3) as well as make the replacement
$\pa_{\t}  \to \pa_{\DP}$. We find
$$~~~~~~~~~~~~{D}_{I i +} { \zeta }_{-~J j}  ~=~ -  C_{i j}  C_{I J}
 {\bar {\cal C}}  ~~~,~~~~~~~~~~
{}~~~~~~~~~~~~~~~~~~~~~~~~~~~~~~~~~~~~~~~~~~~~~~~~~~$$
$$
{}~~~~{\bar D}^{I i }{}_+ { \zeta }_{-~ J j}  ~=~ i  {\cal F} {}_J {}^I  ~
\d_j {}^i ~+~  {\cal F} {}_j {}^i  ~ \d_J{}^I ~~~,~~~~~~~~~~~~~~~~
{}~~~~~~~~~~~~~~~~~~~~~~~~ $$
$$~~~~~~~~~~~~~{D}_{I i +} {\cal C}  ~=~  i 2  C_{i j}  C_{I J}
\pa_{\DP} {\bar {\zeta}_-}{}^{J j}  ~~~,
{}~~~~~~~~~~~~~~~~~~~~~~~~~~~~~~~~~~~~~~~~~~~~~~~~~$$
$$~~~~{\bar D}^{I i }{}_+ {\cal C}  ~=~ 0 ~~~, ~~~~~~~~
{}~~~~~~~~~~~~~~~~~~~~~~~~~~~~~~~~~~~~~~~~~~~~~~~~~~~$$
$$~~~~~~~~~~~~~~~~~~~~~~~{D}_{I i +} {\cal F} {}_j {}^k ~=~   i 2 \left [
\d_i {}^k ( \pa_{\DP} { \zeta }_{-~ I j}) ~-~ \fracm 12 \d_j {}^k
(\pa_{\DP} {\zeta}_{-~I i})   \right ] ~~~,~~~~
{}~~~~~~~~~~~~~~~~~~~~~~~~~~~~~~~~~~~~~~~~~~~~~~$$
\begin{equation}
{}~~~~~~~~~~~~~~~~~~~~~{D}_{I i +} {\cal F} {}_J {}^K ~=~ 2 \left [
\d_I {}^K (\pa_{\DP} { \zeta }_{-~J i}) ~-~ \fracm 12 \d_J {}^K
(\pa_{\DP} {\zeta}_{-~I i})  \right ] ~~~,~~~~~~~~~~~~
{}~~~~~~~~~~~~~~~~~~~~~~~~~~~~~~~~~~~~~~~~~~~~~~~~~~~
\end{equation}
also provides a realization of the (8,0) heterotic supersymmetry
algebra.

{}~~~~In the next section we complete the ``oxidation'' by obtaining
the 2D, N = 8 ultra-multiplet.  The careful reader may at this point
object, ``How can it be that by reducing a 2D, N = 4 model to 1D,
performing a 1D automorphic duality map and then oxidizing back, we
find a 2D, N = 8 theory?'' This almost appears to be magic!  It is
not quite.  The 2D, N = 4 representation from which we started was
an off-shell representation. The 2D, N = 8 representation that we
find after oxidation is an on-shell theory realizing the supersymmetry
algebra,
\begin{equation}
\{ {D}_{I i \a}~,~ {D}_{J j \b} \} ~=~ 0 ~~~,~~~  \{ {D}_{I i \a}~,~
{\bar {D}}^{J j }{}_{\b} \} ~=~ i 2 \d_i {}^j  \d_I {}^J
( \g^c)_{\a \b} \pa_c ~~~.
 \end{equation}
So the original physical plus auxiliary degrees of freedom are converted
via reduction, auto-dualization \& oxidation into purely physical degrees
of freedom afterward. This is the power of 1D automorphic duality!  If
at a later point, we are able to find the off-shell formulation of the
2D ultra-multiplet, then this process can be repeated to derive an
on-shell N = 16 theory.  The construction of the off-shell
ultra-multiplet will require the ``integration'' of the fields of
an UM together with those of a FUM.

\section{The Basic 2D, N = 8 Ultra-multiplet\newline Representation}

{}~~~~In the last section, we saw how 2D, N = 8 ultra-multiplets can
actually be derived by starting from 2D, N = 4 hypermultiplets.
Here we start by giving the simplest ultra-multiplet action
$${\cal L}_{{\rm {UM}}} ~=~ [ ~ \fracm 12 (\pa^a {\bar {\cal G}} )
(\pa_a { {\cal G}} )
{}~+~ \fracm 14 (\pa^a {L} {}_j {}^k )  (\pa_a  {L} {}_k {}^j) ~+~
{}~~~~~~~~~~~~~~~$$
\begin{equation}
{}~~~~~~~~~ \fracm 14 (\pa^a  {R} {}_J {}^K)  (\pa_a  {R} {}_K {}^J)
{}~-~ i {\bar {\varphi}}^{I i \a} (\g^c )_{\a \b} \pa_c
{ \varphi}^{\b}{}_{I i} ~ ]  ~~~,
\end{equation}
which is left invariant under the 2D, N = 8 supersymmetry variations
given by
$$~~~~~~~~~~~~~{D}_{I i \a} {\cal G}  ~=~  i 2  C_{i j}  C_{I J}
{\bar {\varphi}}{}_{\a} {}^{J j}  ~~~,
{}~~~~~~~~~~~~~~~~~~~~~~~~~~~~~~~~~~~~~~~~~~~~~~~~~~$$
$$~~~~{{\bar D}_{\a}}^{I i } {\cal G}  ~=~ 0 ~~~, ~~~~~~
{}~~~~~~~~~~~~~~~~~~~~~~~~~~~~~~~~~~~~~~~~~~~~~~~~~~$$
$$~~~~~~~~~~~~~~~~~~~~~{D}_{I i \a} {L} {}_j {}^k ~=~   i 2 \left [
\d_i {}^k { \varphi }_{I j \a} ~-~ \fracm 12 \d_j {}^k {\varphi}_{I i \a}
\right ] ~~~,~~~~
{}~~~~~~~~~~~~~~~~~~~~~~~~~~~~~~~~~~~~~~~~~~~~~~~~$$
$$~~~~~~~~~~~~~~~~~~~~~{D}_{I i \a} {R} {}_J {}^K ~=~    2 \left [
\d_I {}^K { \varphi }_{J i \a} ~-~ \fracm 12 \d_J {}^K {\varphi}_{I i \a}
\right ] ~~~,~~~~~~~~
{}~~~~~~~~~~~~~~~~~~~~~~~~~~~~~~~~~~~~~~~~~~~~~~~~$$
$$~~~~~~~~~~~~~{D}_{I i \a} { \varphi }_{J j \b}  ~=~ -  C_{i j}  C_{I J}
(\g^c)_{\a \b} (\pa_{c} {\bar {\cal G}} ) ~~~,~~~
{}~~~~~~~~~~~~~~~~~~~~~~~~~~~~~~~~~~~~~~~~~~~$$
\begin{equation}
{}~~~~{\bar D}^{I i \a} { \varphi }_{J j \b}  ~=~ i (\g^c)_{\a \b}
( \pa_{c}  {R} {}_J {}^I ) ~ \d_j {}^i ~+~ (\g^c)_{\a \b}
( \pa_{c} {L} {}_j {}^i ) ~ \d_J{}^I ~~~.~~~~
{}~~~~~~~~~ \end{equation}
One of the most interesting features of the ultra-multiplet is the
group of automorphism that it realizes on the 8 supersymmetry
generators.  The group turns out to be $SU(2) \otimes SU(2) \otimes
U(1) $.  This non-semisimple group is much smaller than the expected
$SO(8)$ normally assumed to appear in a 2D, N = 8 superconformal theory.
As can be seen from the action, this theory is clearly scale invariant.
In fact, the existence of the UM and FUM theories, suggests the existence
of an (8,0) (as well as N = 8) 2D supergravity multiplet with only
seven gauge fields gauging $SU(2) \otimes SU(2) \otimes U(1)$.

\section{Parity Twists of the Ultra-Multiplet Theory}

{}~~~~Sometime ago, the concepts of variant representations \cite{G2}
and twisted multiplets \cite{GaHR} were introduced introduced.
These are useful to recall, because they allow us to use the
basic ultra-multiplet to derive additional representations of
the N = 8 supersymmetry.  The use of a parity twist is a useful
way to find these. The idea is simple.  Given a representation of
2D supersymmetry, it is possible to find a new and distinct
representation by replacing scalar spin-0 fields by pseudo-scalar
spin-0 fields.  Thus, there are variants ultra-multiplets that
contain one, two, three and four pseudo-scalars (any more than
this is equivalent to one of these cases).  We will refer to these
as the twisted ultra-multiplets I thru IV (i.e. TUM-I, TUM-II,
TUM-III and TUM-IV).

{}~~~~We begin our discussion by considering the TUM-I theory.  The parity
twist is incorporated into this model by defining its supersymmetry
variations as,
$$~~~~~~~~~~~~~{D}_{I i \a} {\tilde {\cal A}}  ~=~  i 2  C_{i j}  C_{I J}
{\bar {\rho}}{}_{\a} {}^{J j}  ~~~,
{}~~~~~~~~~~~~~~~~~~~~~~~~~~~~~~~~~~~~~~~~~~~~~~~~~~$$
$$~~~~~~~~~~~~~{D}_{I i \a}  {\tilde {\cal B}}  ~=~  - 2  C_{i j}  C_{I J}
( \g^3)_{\a} {}^{\d}{\bar {\rho}}{}_{\d} {}^{J j}
 ~~~, ~~~~~~
{}~~~~~~~~~~~~~~~~~~~~~~~~~~~~~~~~~~~$$
$$~~~~~~~~~~~~~~~~~~~~~{D}_{I i \a} {\tilde {L}} {}_j {}^k ~=~   i 2 \left [
\d_i {}^k { \rho }_{I j \a} ~-~ \fracm 12 \d_j {}^k {\rho}_{I i \a}
\right ] ~~~,~~~~
{}~~~~~~~~~~~~~~~~~~~~~~~~~~~~~~~~~~~~~~~~~~~~~~~~$$
$$~~~~~~~~~~~~~~~~~~~~~{D}_{I i \a} {\tilde {R}} {}_J {}^K ~=~    2 \left [
\d_I {}^K { \rho }_{J i \a} ~-~ \fracm 12 \d_J {}^K {\rho}_{I i \a}
\right ] ~~~,~~~~~~~~
{}~~~~~~~~~~~~~~~~~~~~~~~~~~~~~~~~~~~~~~~~~~~~~~~~$$
$$~~~~~~~~~~~~~~{D}_{I i \a} { \rho }_{J j \b}  ~=~ -  C_{i j}  C_{I J} [~
(\g^c)_{\a \b} (\pa_{c} {\tilde {\cal A}} )
{}~-~ i (\g^3 \g^c)_{\a \b} (\pa_{c} {\tilde {\cal B}} ) ~] ~~~,~~~
{}~~~~~~~~~~~~~~$$
\begin{equation}
{}~~~~~{\bar D}^{I i \a} { \rho }_{J j \b}  ~=~ i (\g^c)_{\a \b}
( \pa_{c} {\tilde {R}} {}_J {}^I ) ~ \d_j {}^i ~+~ (\g^c)_{\a \b}
( \pa_{c} {\tilde {L}} {}_j {}^i ) ~ \d_J{}^I ~~~.~~~~
{}~~~~~~~~ \end{equation}
Above the component field ${\cal B}$ is the pseudo-scalar that
replaces a scalar in the basic ultra-multiplet.  The action for the
multiplet is exactly the same (in form) as that for the basic
ultra-multiplet.  One of the most interesting aspects of the TUM-I
model is that it is related to a 2D, N = 8 vector multiplet.  This is
 most clearly seen by writing the commutator algebra for the 2D, N
= 8 gauge U(1) covariant derivative.
$$ [~ \nabla_{i I \a} ~,~ \nabla_{j J \b} ~ \} ~=~ - 4 g C_{\a \b}
[~  C_{I J } {\tilde {L}} {}_i {}^k C_{k j} ~-~ i
 C_{i j } {\tilde {R}} {}_I {}^K C_{K J} ~]  ~~~, ~~~~~~~~~~~~~~~~ $$
$$ [~ \nabla_{i I \a} ~,~{\bar \nabla}^{j J}{}_{ \b} ~ \} ~=~ i 2 \d_i {}^j
 \d_I {}^J (\g^c)_{\a \b} \nabla_c ~+~ 2 g \d_i {}^j \d_I {}^J
[~ C_{\a \b} {\tilde {\cal A}} ~+~ i (\g^3)_{\a \b} {\tilde {\cal B}} ~] ~~~,
$$
$$ [~ \nabla_{i I \a} ~,~{\nabla}_{ b} ~ \} ~=~ g  (\g^c)_{\a}{}^{ \g}
{\bar W}_{i I \g}   ~~~, ~~~~~~~~~~~~~~~~~~~~~~~~~~~~~~~~~~~~~~~~~
{}~~~~~~ $$
\begin{equation}
[~ \nabla_{a} ~,~{\nabla}_{ b} ~ \} ~=~ - ig  \e_{ a b} {\cal W}
 ~~~. ~~~~~~~~~~~~~~~~~~~~~~~~~~~~~~~~~~~~~~~~~~~~~~~~~
\end{equation}
The Bianchi Identities that follow from these equations have a
solution that is closely related to those in (9.)\footnote{
It can be observed that these results may be derived by applying
simple dimensional compactification to 4D, N = 4 Yang-Mills
theory.  This provides an interesting and independent confirmation
of the existence of the ultra-multiplet.}.  We need only identify
${\bar W}_{i I \a} = - \fracm 12 C_{i j} C_{I J} {\bar
\rho}^{j J }{}_{\a}$ and to slightly modify one of our previous
results to,
\begin{equation}
{D}_{I i \a} { \rho }_{J j \b}  ~=~  C_{i j}  C_{I J}
(\g^3 )_{\a \b} {\cal W} - C_{i j}  C_{I J} [~
(\g^c)_{\a \b} (\pa_{c} {\tilde {\cal A}} )
{}~-~ i (\g^3 \g^c)_{\a \b} (\pa_{c} {\tilde {\cal B}} ) ~] ~~~.
\end{equation}
This result is the usual one that follows in a 2D supersymmetric
theory when one compares a scalar multiplet to a vector multiplet.
(In an off-shell formulation of the TUM-I model, ${\cal W}$ is
replaced by an auxiliary field. This is the beginning of the
off-shell formulation of the TUM-I theory.)  A final point of
interest regarding this form of the ultra-multiplet is that this
version can be ``oxidized'' all the way back to 4D where it
can be recognized as the 4D, N = 4 Yang-Mills theory.

{}~~~~The next ultra-multiplet is the TUM-II theory which possesses
two pseudo-scalars among its fields. Its supersymmetry variations
are given by
$$~~~~~~~~~~~~~~~~~~~~{D}_{I i \a} {\cal H}  ~=~  i 2  C_{i j}  C_{I J}
 (\g^3)_{\a}{}^{ \b} {\bar {\l}}{}_{\b} {}^{J j}  ~~~,
{}~~~~~~~~~~~~~~~~~~~~~~~~~~~~~~~~~~~~~~~~~~~~~~~~~~$$
$$~~~~~{{\bar D}_{\a}}^{I i } {\cal H}  ~=~ 0 ~~~, ~~~~~~
{}~~~~~~~~~~~~~~~~~~~~~~~~~~~~~~~~~~~~~~~~~~~~~~~~~~$$
$$~~~~~~~~~~~~~~~~~~~~~{D}_{I i \a} {\cal B} {}_j {}^k ~=~   i 2 \left [
\d_i {}^k { \l }_{I j \a} ~-~ \fracm 12 \d_j {}^k {\l}_{I i \a}
\right ] ~~~,~~~~
{}~~~~~~~~~~~~~~~~~~~~~~~~~~~~~~~~~~~~~~~~~~~~~~~~$$
$$~~~~~~~~~~~~~~~~~~~~~{D}_{I i \a} {\cal A} {}_J {}^K ~=~    2 \left [
\d_I {}^K { \l }_{J i \a} ~-~ \fracm 12 \d_J {}^K {\l}_{I i \a}
\right ] ~~~,~~~~~~~~
{}~~~~~~~~~~~~~~~~~~~~~~~~~~~~~~~~~~~~~~~~~~~~~~~~$$
$$~~~~~~~~~~~~~~~~{D}_{I i \a} { \l }_{J j \b}  ~=~ -  C_{i j}  C_{I J}
(\g^3 \g^c)_{\a \b} (\pa_{c} {\bar {\cal H}} ) ~~~,~~~
{}~~~~~~~~~~~~~~~~~~~~~~~~~~~~~~~~~~~~~~~~~~~$$
\begin{equation}
{}~~~~~~~~~~~{\bar D}^{I i \a} { \l }_{J j \b}  ~=~ i (\g^c)_{\a \b}
( \pa_{c}  {\cal A} {}_J {}^I ) ~ \d_j {}^i ~+~  (\g^c)_{\a \b}
( \pa_{c} {\cal B} {}_j {}^i ) ~ \d_J{}^I ~~~.~~~~
{}~~~~~~~~~ \end{equation}

{}~~~~Continuing along the same lines, there is the TUM-III theory
containing three pseudo-scalars in its spectrum. Here the
supersymmetry variations are defined by,
$$~~~~~~~~~~~~~{D}_{I i \a} {\cal M}  ~=~  i 2  C_{i j}  C_{I J}
 {\bar {\Phi}}{}_{\a} {}^{J j}  ~~~,
{}~~~~~~~~~~~~~~~~~~~~~~~~~~~~~~~~~~~~~~~~~~~~~~~~~~$$
$$~~~~{{\bar D}_{\a}}^{I i } {\cal M}  ~=~ 0 ~~~, ~~~~~~
{}~~~~~~~~~~~~~~~~~~~~~~~~~~~~~~~~~~~~~~~~~~~~~~~~~~$$
$$~~~~~~~~~~~~~~~~~~~~~{D}_{I i \a} {X} {}_j {}^k ~=~   i 2 \left [
\d_i {}^k { \Phi }_{I j \a} ~-~ \fracm 12 \d_j {}^k {\Phi}_{I i \a}
\right ] ~~~,~~~~
{}~~~~~~~~~~~~~~~~~~~~~~~~~~~~~~~~~~~~~~~~~~~~~~~~$$
$$~~~~~~~~~~~~~~~~~~~~~~{D}_{I i \a} {Y} {}_J {}^K ~=~ 2 (\g^3)_{\a}{}^{\b}
\left [
\d_I {}^K { \Phi }_{J i \b} ~-~ \fracm 12 \d_J {}^K {\Phi}_{I i \b}
\right ] ~~~,~~~~~~~~
{}~~~~~~~~~~~~~~~~~~~~~~~~~~~~~~~~~~~~~~~~~~~~~~~~$$
$$~~~~~~~~~~~~~~~~{D}_{I i \a} { \Phi }_{J j \b}  ~=~ -  C_{i j}  C_{I J}
(\g^c)_{\a \b} (\pa_{c} {\bar {\cal M}} ) ~~~,~~~
{}~~~~~~~~~~~~~~~~~~~~~~~~~~~~~~~~~~~~~~~~~~~$$
\begin{equation}
{}~~~~~~~~~~{\bar D}^{I i \a} { \Phi }_{J j \b}  ~=~ i  (\g^3 \g^c)_{\a \b}
( \pa_{c}  {Y} {}_J {}^I ) ~ \d_j {}^i ~+~ (\g^c)_{\a \b}
( \pa_{c} {X} {}_j {}^i ) ~ \d_J{}^I ~~~.~~~~
{}~~~~~~~ \end{equation}

{}~~~~Finally, there is the TUM-IV theory containing four pseudo-scalars
in its spectrum. Analogous to the previous versions of this
theory, the supersymmetry variations are given by,
$$~~~~~~~{D}_{I i \a} {\tilde {\cal M}}  ~=~  i 2  C_{i j}  C_{I J}
{\bar {\L}}{}_{\a} {}^{J j}  ~~~, ~~
{}~~~~~~~~~~~~~~~~~~~~~~~~~~~~~~~~~~~~~~~~~~~~~~~~~~$$
$$~~~~~~{D}_{I i \a}  {\tilde {\cal N}}  ~=~  - 2  C_{i j}  C_{I J}
( \g^3)_{\a} {}^{\d}{\bar {\L}}{}_{\d} {}^{J j}
 ~~~, ~~~~~~~
{}~~~~~~~~~~~~~~~~~~~~~~~~~~~~~~~~~~~$$
$$~~~~~~~~~~~~~~~~~~{D}_{I i \a} {U} {}_j {}^k ~=~   i 2
( \g^3)_{\a} {}^{\d} \left [
\d_i {}^k {\L }_{I j \d} ~-~ \fracm 12 \d_j {}^k {\L}_{I i \d}
\right ] ~~~,~~~~
{}~~~~~~~~~~~~~~~~~~~~~~~~~~~~~~~~~~~~~~~~~~~~~~~~$$
$$~~~~~~~~~~~~~~~{D}_{I i \a} {V} {}_J {}^K ~=~    2 \left [
\d_I {}^K {\L }_{J i \a} ~-~ \fracm 12 \d_J {}^K {\L}_{I i \a}
\right ] ~~~,~~~~~~~~
{}~~~~~~~~~~~~~~~~~~~~~~~~~~~~~~~~~~~~~~~~~~~~~~~~$$
$$~~~~~~~~{D}_{I i \a} {\L }_{J j \b}  ~=~ -  C_{i j}  C_{I J} [~
(\g^c)_{\a \b} (\pa_{c} {\tilde {\cal M}} )
{}~-~ i (\g^3 \g^c)_{\a \b} (\pa_{c} {\tilde {\cal N}} ) ~] ~~~,~~~
{}~~~~~~~~~~~~~~$$
\begin{equation}
{}~~~~~{\bar D}^{I i \a} {\L }_{J j \b}  ~=~ i (\g^c)_{\a \b}
( \pa_{c} {V} {}_J {}^I ) ~ \d_j {}^i ~+~ (\g^3 \g^c)_{\a \b}
( \pa_{c} U_j {}^i ) ~ \d_J{}^I ~~~.~~~~
{}~~~~~~~~ \end{equation}

{}~~~~The form of the action for all of the ultra-multiplets is given
by equation (7).

\section{SU(2) $\otimes$ SO(2) Ultra-Multiplets}

{}~~~~The starting point of our discussions was the reduction, dualization
and oxidation of the THM-II, N = 4 theory.  However, we also could
have used the THM-I, N = 4 theory as the starting point! Carrying
out the reduction leads to the intermediate results,
$$\begin{array}{lll}
D_{I i} F  &=& 2  C_{i j} \l_{I} {}^j ~~~, \\

{\bar D}_{I}{}^{ i} F  &=& 0 ~~~, \\

D_{I i} S  &=& -i  {\bar \l}_{I i}  ~~~, \\

D_{I i} P  &=&  (\s^3)_{I} {}^{J} {\bar \l }_{J i} ~~~, \\

D_{I i} \l^{J}{}^{j} &=&  \d_i {}^j \left [~ \d_{I J}
(\pa_{\t} S) ~+~ i(\s^3)_{I J } (\pa_{\t} P)~\right ]  \\
& &- i ~   \left [ ~ \fracm 12 \d_i {}^j (\s^1)_{I J} (\pa_{\t}
\varphi) ~-~ 2  (\s^2)_{I J} (\pa_{\t} \varphi_i {}^j ) ~\right ] ~~, \\

{\bar D}_{I}{}^{i} \l_J {}^{j} &=& i C^{i j}\d_{I J } (\pa_{\t} F) ~~, \\

D_{\a i} \varphi &=& - 2  (\s^1)_{I} {}^{J} {\bar \l }_{J i}  ~~, \\
\end{array}    $$
\begin{equation}
D_{I i} \varphi_j {}^k  = {~~}- (\d_j {}^l \d_i {}^k - \frac 12
\d_j {}^k \d_i {}^l ) (\s^2)_{I} {}^{J} {\bar \l
}_{J l}  ~~, {~~~~~}
\end{equation}
As can be seen this 1D theory only has only $SO(2) \otimes SU(2) \otimes
U(1) $ symmetry.  Its oxidation back to 2D retains this structure.  This
is the beginning of a whole set of similar such theories.   But all of
these theories are related by a redefinition to the previous discussed
theories.  In particular, one need only perform the redefinition
$D_{I i \a} \to (\s^2)_I {}^J  D_{J i \a} $.

\section{Summary and Conclusion}

{}~~~~We have seen that rigid 2D, N = 8 representations are very abundant.
The ultra-multiplet, in all of its guises, manifest a very small
group of automorphisms on the supersymmetry derivatives (typically
only $SU(2) \otimes SU(2) \otimes U(1) $ or $SO(2) \otimes SU(2) \otimes
U(1)$. It is trivially the case that the foremost of these can also
be regarded as $SO(4) \otimes U(1)$ or $SO(4) \otimes SO(2) $ groups.
All of these are much smaller than the expected $SO(8)$ of the
known 2D, N = 8 superconformal theories. Furthermore the rigid actions
for all of these models are scale invariant.  The real remaining
challenge is to find out whether there exist 2D, N = 8 conformal
supergravity theories that can be coupled to ultra-multiplets. None
of the standard constructions associated with conformal supergroups
seem compatible with ultra-multiplets!  It is just possible that
presently unknown N = 8 string-like theories may be waiting to
be discovered.

$${~~~~}$$

$${~~~~}$$

$${~~~~}$$

$${~~~~}$$

{\it Acknowledgment}

One of the authors (S.J.G.) wishes to thank the organizers of the ``
The Quantum Field Theory \indent and Gravity'' Conference held in
Tomsk, Russia during August 22-27, 1994 where \indent this
investigation was initiated.

\noindent{{\bf {Appendix: $SO(4) \otimes SO(2)$ Formulation of the
Ultra-multiplet}}}

{}~~~~In this appendix, we wish to give an alternate formulation of the
ultra-multiplet. We wish to take advantage of the fact that
$SU(2) \otimes SU(2) \otimes U(1)$ is equivalent to $SO(4) \otimes
U(1)$  This implies that the spinor derivative may be given in
the form $D_{i \a}$ here the $i$-index takes on four values (i.e.
vector indices in $SO(4)$). These derivatives are still complex
so a rigid phase rotation may act to realize a U(1) (SO(2)) symmetry
upon them.  The component fields of the basic ultra-multiplet
can be expressed as ${\cal G}$, $\varphi_{i \a}$,
$L_{\hat a}$ and $R_{\hat a}$. The supersymmetry variations
of the 1D theory take the form,
$$~~~~~~~~~~~~~~~~~{D}_i {\cal G}  ~=~  i 2 \d_{i j}
{\bar {\varphi}}^{j}  ~~~,
{}~~~~~~~~~~~~~~~~~~~~~~~~~~~~~~~~~~~~~~~~~~~~~~~~~~$$
$$~~~~~~~~~~~~~~~{\bar D}^{i } {\cal G}  ~=~ 0 ~~~, ~~~~~~
{}~~~~~~~~~~~~~~~~~~~~~~~~~~~~~~~~~~~~~~~~~~~~~~~~~~$$
$$~~~~~~~~~~~~~~~~~~~~~{D}_{i } L_{\hat a} ~=~   i 2
(\a_{\hat a})_i {}^j { \varphi }_{ j} ~~~,~~~~
{}~~~~~~~~~~~~~~~~~~~~~~~~~~~~~~~~~~~~~~~~~~~~~~~~$$
$$~~~~~~~~~~~~~~~~~~~~~~~{D}_{i} R_{\hat a} ~=~    2 (\b_{\hat a})_i {}^j
{ \varphi }_{ j} ~~~,~~~~~~~~
{}~~~~~~~~~~~~~~~~~~~~~~~~~~~~~~~~~~~~~~~~~~~~~~~~$$
$$~~~~~~~~~~~~~~~~~~~{D}_{i } { \varphi }_{j}  ~=~ -  \d_{i j}
(\pa_{\t} {\bar {\cal G}} ) ~~~,~~~
{}~~~~~~~~~~~~~~~~~~~~~~~~~~~~~~~~~~~~~~~~~~~~~~~~$$
$$
{}~~~~~~~~~~~~~~~~{\bar D}^{i } { \varphi }_{j}  ~=~ (\a^{\hat a})_j {}^i
( \pa_{\t}  L_{\hat a} ) ~+~  i  (\b^{\hat a})_j {}^i
( \pa_{\t}  R_{\hat a}  ) ~
 ~ ~~~.~~~~~~~~~~~~~ \eqno(A.1) $$
In these expressions, the quantities $ (\a^{\hat a})_i {}^j$
and $(\b^{\hat a})_i {}^j$ represent two commuting sets of real,
four by four, antisymmetric $SU(2)$ matrices.  Taken together these
six matrices represent the generators of $SO(4)$.  These quantities
are well known in the physics literature \cite{AB}. The Lagrangian
can be written concisely as
$${\cal L}_{{\rm {UM}}} ~=~ [ ~ \fracm 14 (\pa_{\t} {{\bar S}} {}_j
{}^k )  (\pa_{\t}  {S} {}_k {}^j)  ~-~ i {\bar {\varphi}}^{ i }
\pa_{\t } { \varphi}_{ i} ~ ]  ~~~,
\eqno(A.2) $$
where $S_i {}^j \equiv ~ \d_i {}^j {\cal G}  ~+~ (\a^{\hat a})_i {}^j
L_{\hat a}  ~+~  i  (\b^{\hat a})_i {}^j R_{\hat a} $.

\newpage

\end{document}